\documentclass[runningheads]{llncs}
\usepackage{graphicx}
\usepackage{hyperref}
\usepackage{listings}
\usepackage{xcolor}

\definecolor{codegreen}{rgb}{0,0.6,0}
\definecolor{codegray}{rgb}{0.5,0.5,0.5}
\definecolor{codepurple}{rgb}{0.58,0,0.82}
\definecolor{backcolour}{rgb}{0.95,0.95,0.92}

\lstdefinestyle{mystyle}{
    backgroundcolor=\color{backcolour},   
    commentstyle=\color{codegreen},
    keywordstyle=\color{magenta},
    numberstyle=\tiny\color{codegray},
    stringstyle=\color{codepurple},
    basicstyle=\ttfamily\footnotesize,
    breakatwhitespace=false,         
    breaklines=true,                 
    captionpos=b,                    
    keepspaces=true,                 
    numbers=left,                    
    numbersep=5pt,                  
    showspaces=false,                
    showstringspaces=false,
    showtabs=false,                  
    tabsize=2
}

\lstdefinestyle{yaml}{
    basicstyle=\ttfamily\small,
    keywordstyle=\color{blue},
    breaklines=true,
    showstringspaces=false,
    identifierstyle=\color{black},
    stringstyle=\color{red},
    commentstyle=\color{gray},
}

\lstdefinelanguage{yaml}{
    morekeywords={true,false,null,y,n},
    sensitive=false,
    morecomment=[l]{\#},
    morestring=[b]",
    morestring=[b]',
    style=yaml
}

\begin{document}
\title{Aorta Segmentation from 3D CT in MICCAI SEG.A. 2023 Challenge.}
%
%

\author{Andriy Myronenko \and 
Dong Yang \and
Yufan He \and
Daguang Xu 
}

\authorrunning{A. Myronenko et al.}
%
\institute{NVIDIA \\
\email{amyronenko@nvidia.com}}

\maketitle

\begin{abstract}
Aorta provides the main blood supply of the body. Screening of aorta with imaging helps for early aortic disease detection and monitoring.  In this work, we describe our solution to the Segmentation of the Aorta (SEG.A.23\footnote{https://multicenteraorta.grand-challenge.org/multicenteraorta/}) from 3D CT challenge. We use automated segmentation method Auto3DSeg\footnote{https://monai.io/apps/auto3dseg} available in MONAI\footnote{https://github.com/Project-MONAI/MONAI}.  Our solution achieves an average Dice score of 0.920 and 95th percentile of the Hausdorff Distance (HD95) of 6.013,  which ranks first  and wins the SEG.A. 2023 challenge.

\end{abstract}


\section{Introduction}

\begin{figure}[t]
    \centering
    \includegraphics[width=0.425\textwidth]{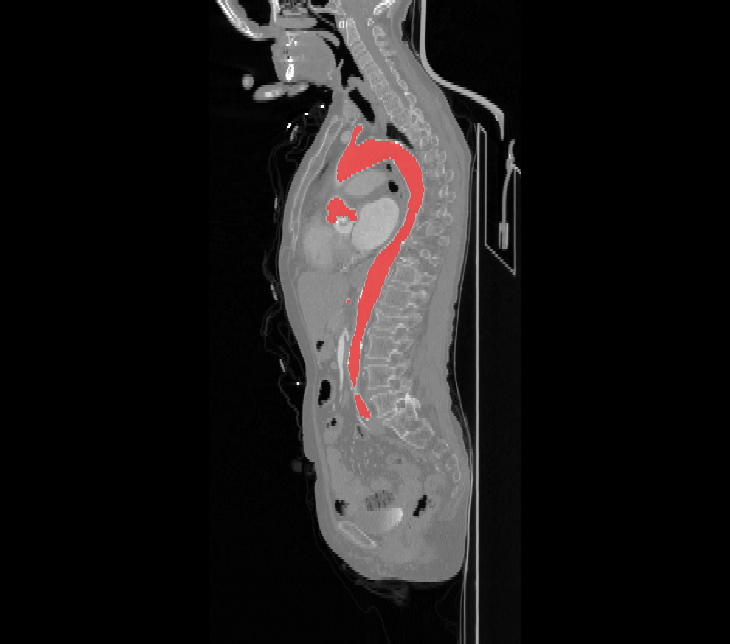}
    \includegraphics[width=0.45\textwidth]{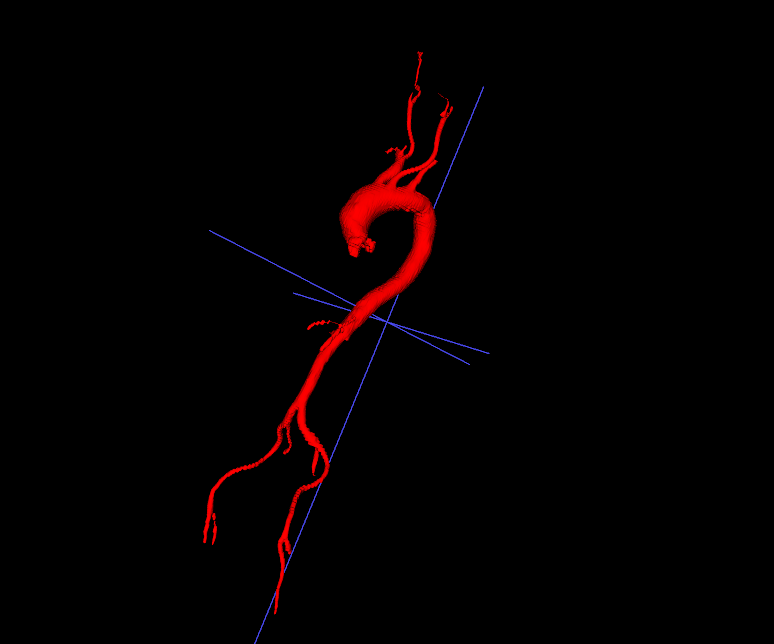}

    \caption{SEG.A.23 data example of a sagittal 3D CT slice. The provided aorta segmentation (in red) is also shown as a 3D rendering.}
    \label{fig:example}
\end{figure}

The aorta serves as the principal artery within the human body. To effectively monitor patients with aortic diseases, it is necessary to regularly examine the vessels for any signs of disease progression~\cite{PEPE2020101773,jin2023aibased}. Computed tomography angiography (CTA) stands as the conventional imaging technique for clinical evaluation, offering a comprehensive visual of the aortic vessel tree (AVT)\cite{RADL2022107801}. Segmentation of the Aorta (SEG.A.23) 2023 challenge aims to evaluate and compare AI techniques for the best aorta segmentation. The challenge provides the participants with a training set of AVTs and the corresponding manual segmentations from three institutions. The participants are expected to design algorithms for automatic AVT segmentation (see Fig.~\ref{fig:example}). All the proposed methods are evaluated based on a hidden test set from a fourth institution using Dice Similarity Score (DSC) and 95th percentile of the Hausdorff Distance (HD95). An optional sub-tasks also asks to create a reconstructed 3D surface geometry, to be evaluated for meshing quality.

\section{Methods}

We implemented our approach with MONAI~\cite{monai} using Auto3DSeg open source project. Auto3DSeg is an automated solution for 3D medical image segmentation, utilizing open source components in MONAI, offering both beginner and advanced researchers the means to effectively develop and deploy high-performing segmentation algorithms.

The minimal user input to run Auto3DSeg for SEG.A.23 datasets, is 
\begin{lstlisting}[language=bash]
#!/bin/bash
python -m monai.apps.auto3dseg AutoRunner run \
    --input=./input.yaml --algos=segresnet
\end{lstlisting}

where a user provided input configuration (input.yaml) including only 3 lines:

\begin{lstlisting}[language=yaml]
# This is the YAML file "input.yaml"
modality: CT
datalist: ./dataset.json
dataroot: /data/seg.a23
\end{lstlisting}

When running this command,  Auto3DSeg will analyze the dataset, generate hyperparameter configurations for several supported algorithms, train them, and produce inference and ensemble.  The system will automatically scale to all available GPUs  and also supports multi-node training.  The 3 minimum user options (in input.yaml) are data modality (CT in this case), location of the downloaded SEG.A. dataset (dataroot), and the list of input filenames with an associated fold number (dataset.json). We generate the 5-fold split assignments randomly.

Currently, the default Auto3DSeg setting trains three 3D segmentation algorithms: SegResNet~\cite{myronenko20183d}, DiNTS~\cite{he2021DiNTS} and SwinUNETR~\cite{hatamizadeh2021swin,tang2022self} with their unique training recipes. SegResNet and DiNTS are convolutional neural network (CNN) based architectures, whereas SwinUNETR is based on transformers. Here we used only SegResNet.



In the final prediction, we ensemble 15 best model checkpoints of SegResNet  (5-folds trained 3 times). 


\subsection{SegResNet}

\begin{figure}[t]
    \centering
    \includegraphics[width=0.8\textwidth]{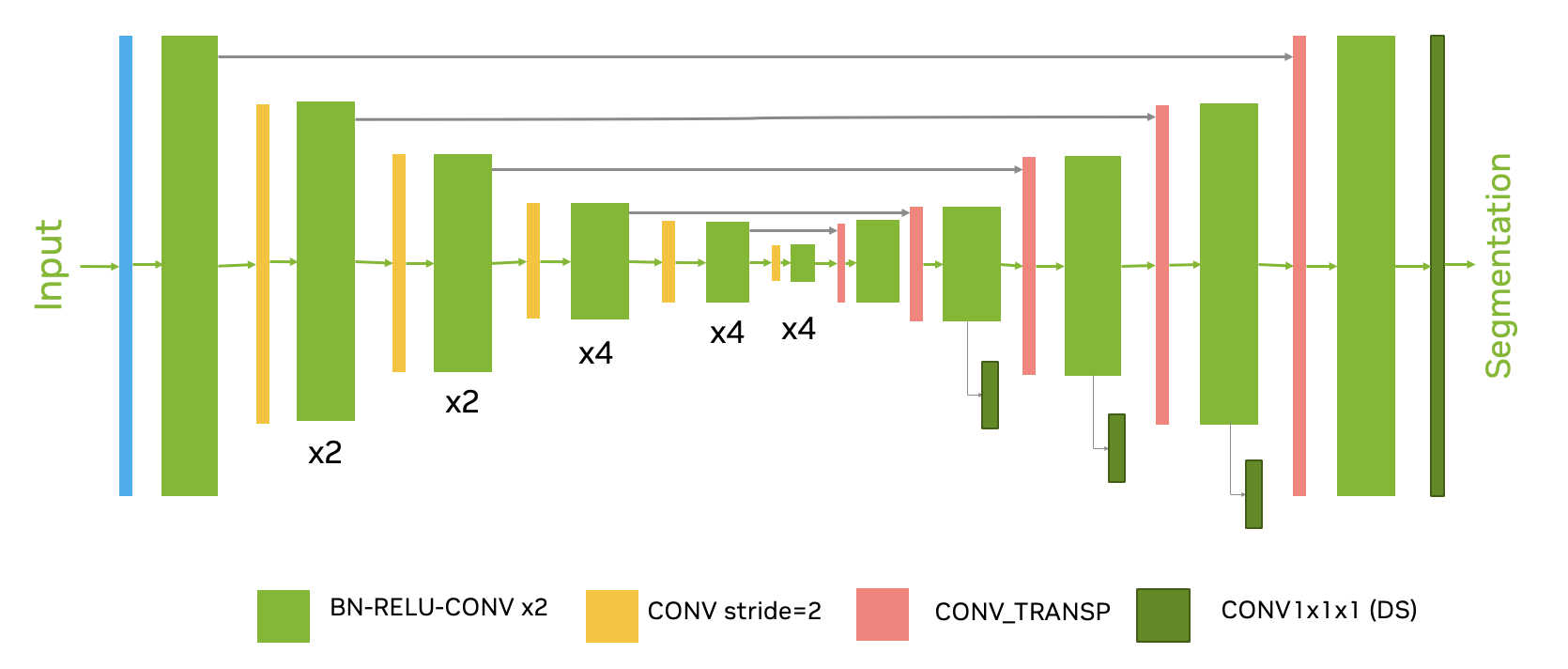}

    \caption{SegResNet network configuration. The network uses repeated ResNet blocks with batch normalization and deep supervision}
    \label{fig:net}
\end{figure}

SegResNet\footnote{https://docs.monai.io/en/stable/networks.html} is an encode-decoder based semantic segmentation network based on~\cite{myronenko20183d} It is a U-Net based convolutional neural network with deep supervision (see Figure~\ref{fig:net}).

The encoder part uses residual network blocks, and includes 5 stages of 1, 2, 2, 4, 4  blocks respectively. It follows a common CNN approach to downsize image dimensions by 2 progressively and simultaneously increase feature size by 2.  All convolutions are $3 \times 3 \times 3$ with an initial number of filters equal to 32.  The decoder structure is similar to the encoder one, but with a single block per each spatial level. Each decoder level begins with upsizing with transposed convolution: reducing the number of features by a factor of 2  and doubling the spatial dimension, followed by the addition of encoder output of the equivalent spatial level. The number of levels and the region size is automatically configured.  We use spatial augmentation including random affine transforming and flipping in all axes, random intensity scaling, shift, noise and blurring.



\subsection{Optimization}
The models in our solution were trained on an 8-GPU NVIDIA V100 (32 GB) machine. We use the AdamW optimizer with an initial learning rate of $2e^{-4}$ and decrease it to zero at the end of the final epoch using the Cosine annealing scheduler. The batch size is 1 per GPU (effective batch size 8), and weight decay regularization is $1e^{-5}$.

We trained each of the 5-folds 3 times, and ensemble the final 15 models. Then we used the combined dice and focal loss, random cropping of $288 \times 288 \times 288$, and resampled all input images to $0.7 \times 0.7 \times 1.0 \mathrm{mm}^3$ common resolution.  
Since one of the goals of SEG.A.23 is to evaluate method robustness to various intensity difference in data from different institutions, we decided to implement an adaptive input 3D CT image normalization. Specifically the Aorta segmentation result from the ensemble of the first 5 model checkpoints (5-folds) was used to detect the CT intensity range within the aorta foreground region. The 5th and 95th percentile intensity bounds of this region were used to re-scale 3D CT intensity globally from this range to 0..1 with soft clipping.  And the remaining 10 model inferences were produced on this re-normalized 3D CT image.  For the first 5 models, we normalize input images to zero mean, unit standard deviation of the intensity globally.  This deviates from the default Auto3DSeg approach to normalize CT images from a foreground range to 0..1. The reasons for this choice was that some SEG.A.23 CT images were in proper Hounsfield units, and some were saved only in positive values probably for compression reasons (which is a non-standard input CT format). Generally, CT scanners are calibrated to produce a proper Hounsfield unit, so in this case our adaptive normalization may not be necessary.  This intensity re-normalization approach showed only a slight advantage in our cross-validation experiments, but we decided to keep it.

\section{Results}

Based on our random 5-fold split, the average dice scores per fold are shown in Table~\ref{tab:result}.  

\begin{table}[h!]
    \centering
    \begin{tabular}{| c | c | c | c | c | c |}
        \hline
        {\textbf{Fold 1}} & {\textbf{Fold 2}} & {\textbf{Fold 3}} & {\textbf{Fold 4}} & {\textbf{Fold 5}} & {\textbf{Average}} \\
        \hline
        0.9298 & 0.9432	& 0.9478 & 0.9401& 0.9547 & 0.94312 \\
        \hline
    \end{tabular}
    \caption{Average Dice results of a single model based on our 5-fold data split.}
    \label{tab:result}
\end{table}

On the final hidden challenge dataset, our submission achieved an average Dice score of 0.920 and HD95 of 6.013. SEG.A.23 challenge used a combined metric to rank the methods (accounting for  various statistics of Dice and Hausdorff distance) to comprehensively evaluate both method accuracy and robustness.  Our submission ranked first on the final test set\footnote{https://multicenteraorta.grand-challenge.org/main-task-final-leaderboard/}.

\section{Conclusion}
We described our winning solution to SEG.A. 2023 challenge using Auto3DSeg from MONAI.
Our final submission is an ensemble of 15 SegResNet models. Our solution achieves an average dice of 0.920 and HD95 of 6.013, which ranks first on the SEG.A. 2023 final leaderboard.

\bibliographystyle{splncs04}
\bibliography{paper}

\end{document}